\documentclass[english,aps,prb,reprint,superscriptaddress]{revtex4-1}
\usepackage[T1]{fontenc}
\usepackage[latin9]{inputenc}
\usepackage[letterpaper]{geometry}
\usepackage{color}
\usepackage{float}
\usepackage{bm}
\usepackage{graphicx}

\makeatletter
\@ifundefined{showcaptionsetup}{}{%
 \PassOptionsToPackage{caption=false}{subfig}}
\usepackage{subfig}
\makeatother

\usepackage{babel}
\begin{document}
\global\long\def\hc{\text{h.c.}}

\title{High-Energy Damping by Particle-Hole Excitations in the Spin-Wave Spectrum of Iron-Based Superconductors}

\author{Zhidong Leong}

\affiliation{Department of Physics, University of Illinois at Urbana-Champaign,
Urbana IL 61801, USA}

\author{Wei-Cheng Lee}

\affiliation{Department of Physics, University of Illinois at Urbana-Champaign,
Urbana IL 61801, USA}

\author{Weicheng Lv}

\affiliation{Department of Physics and Astronomy, University of Tennessee, Knoxville,
Tennessee 37996, USA}

\affiliation{Materials Science and Technology Division, Oak Ridge National Laboratory,
Oak Ridge, Tennessee 37831, USA}

\author{Philip Phillips}

\affiliation{Department of Physics, University of Illinois at Urbana-Champaign,
Urbana IL 61801, USA}

\begin{abstract}
Using a degenerate double-exchange model, we investigate the spin
excitation spectra of iron pnictides. The model consists of local
spin moments on each Fe site, as well as itinerant electrons from
the degenerate $d_{xz}$ and $d_{yz}$ orbitals. The local moments
interact with each other through antiferromagnetic $J_{1}$-$J_{2}$
Heisenberg interactions, and they couple to the itinerant electrons
through a ferromagnetic Hund coupling. We employ
the fermionic spinon representation for the local moments and perform
a generalized random-phase approximation calculation on both spinons and itinerant electrons.
We find that in the $\left(\pi,0\right)$ magnetically-ordered state,
the spin-wave excitation at $\left(\pi,\pi\right)$ is pushed to a
higher energy due to the presence of itinerant electrons, which is
consistent with a previous study using the Holstein-Primakoff transformation.
In the paramagnetic state, the particle-hole continuum keeps the collective
spin excitation near $\left(\pi,\pi\right)$ at a higher energy even
without any $C_{4}$ symmetry breaking. The implications for recent
high temperature neutron scattering measurements will be discussed.
\end{abstract}
\maketitle

\section{Introduction}

Goldstone's theorem guarantees that the onset of magnetism with a broken continuous symmetry is always accompanied by a gapless spin-wave spectrum in the vicinity of the ordering wave-vector $Q$.  In the iron-pnictide superconductors, inelastic neutron scattering (INS) measurements have remarkably revealed \cite{McQueeney2008,Zhao2008,Ewings2008,Diallo2009,Zhao2009,Diallo2010,Ewings2011,Harriger2011,Harriger2012,Wang2013b,Zhou2013} that even in the paramagnetic state, a well-formed low-energy feature persists in the spin-wave spectrum in the vicinity of the 
$Q=(\pi,0)$ ordering wave vector of the stripe-like magnetic state.  In addition, the high-energy part of the spectrum in the vicinity of $(\pi,\pi)$ remains virtually unchanged even when the temperature is lowered from the paramagnetic to the ordered antiferromagnetic state.  The apparent temperature-independence of the spin-wave spectrum through the magnetic ordering transition is the subject of this paper.

Although both local-moment Heisenberg spin-exchange\cite{Si2008,Han2009,Schmidt2010,Goswami2011,Wysocki2011,Yu2012}
 and itinerant weakly interacting band\cite{Raghu2008,Graser2009,Brydon2009,Knolle2010,Kaneshita2010,Graser2010,Knolle2011} models
have been proposed to explain magnetism in the pnictides, the experimental data\cite{McQueeney2008,Zhao2008,Ewings2008,Diallo2009,Zhao2009,Diallo2010,Ewings2011,Harriger2011,Harriger2012,Wang2013b,Zhou2013} provide ample evidence that neither picture alone will suffice.  INS experiments\cite{McQueeney2008,Diallo2010}  reveal that the
spin-wave spectrum persists up to 200 meV.  While isotropic $J_1$-$J_2$ Heisenberg models can account for the features near the ordering wave vector, they cannot explain the spectrum in the vicinity of $(\pi,\pi)$.  Physically, what would suffice to account for the $(\pi,\pi)$ region is damping arising from particle-hole excitations\cite{Diallo2010}.  The natural source for such excitations is itinerant electrons. 

Hybrid\cite{Kou2009,Lv2010,Yang2010,You2011,You2013} models consisting of local moments and itinerant electrons have already had much success in explaining the INS data.  Lv et al.\cite{Lv2010} considered the local moments in the standard $J_1$-$J_2$ model, where $J_1$ and $J_2$ are the nearest and next-nearest neighbor exchange interactions, respectively,  and the itinerant electrons of the degenerate $d_{xz}$ and $d_{yz}$ orbitals arising from the conduction electrons. The local moments and itinerant electrons were allowed to interact via a ferromagnetic Hund coupling interaction.  The role of the Hund coupling is two-fold.  First, it produces unfrustrated $(\pi,0)$-striped antiferromagnetism.  Previous fits  of the experimentally measured\cite{Zhao2009}  spin-wave dispersion to a pure $J_1$-$J_2$ model required a sizable anisotropy between the exchange interactions along the $x$ and $y$ axes, with one of the interactions becoming ferromagnetic.  The Hund coupling\cite{Lv2010} provided a natural mechanism to explain the origin of this anisotropy, with the added advantage that the magnetism remains unfrustrated.  The second role played by the Hund interaction\cite{Lv2010} is that it lifted the degeneracy of the  $(\pi,0)$ and $(0,\pi)$ magnetic states, giving rise to a relative maximum in the spin-wave spectrum at $(\pi,\pi)$, in contrast to the minimum seen in local moment models\cite{Zhao2009}.  Alternatively, the anisotropy can also be derived within a purely local-moment model with a bi-quadratic coupling between nearest neighbors \cite{Wysocki2011}.
In the paramagnetic state, this model also exhibits features \cite{Goswami2011,Yu2012}
consistent with nematicity found in INS experiments \cite{Harriger2011}.
However, this approach cannot explain the high temperature
INS data \cite{Ewings2011,Harriger2012}, where $C_{4}$ symmetry
is preserved.  In fact, despite the success of  the double-exchange model in generating unfrustrated magnetism, it has not been applied to the paramagnetic high-temperature state.  

In this paper, we use the degenerate double exchange model in
Ref. \onlinecite{Lv2010} to investigate the spin excitation spectra
of iron pnictides in the paramagnetic state. The model consists of local spin moments on each
Fe site, as well as itinerant electrons from the degenerate $d_{xz}$
and $d_{yz}$ orbitals. The local moments interact with each other
through antiferromagnetic $J_{1}$-$J_{2}$ Heisenberg interactions,
and they couple to the itinerant electrons through a ferromagnetic
Hund coupling. Such a local-itinerant model can be motivated by
considering the dual role of $d$ electrons \cite{Gor'kov2013}. Only
$d_{xz}$ and $d_{yz}$ orbitals are included, because they are the
orbitals that break rotational symmetry in the $x$-$y$ plane.  As a  consequence, 
these orbitals form a
minimal model that can drive the magnetic anisotropy. Unlike previous works \cite{Goswami2011,Wysocki2011,Yu2012,Lv2010},
we represent the local moments as fermions. This representation
provides a unified framework for both the ordered and paramagnetic
states, and it yields the Landau damping in addition to the dispersion.
We  then perform a generalized random-phase approximation calculation
on both spinons and itinerant electrons. We  show that, in the
$\left(\pi,0\right)$-magnetically ordered state, the spin-wave excitation
at $\left(\pi,\pi\right)$ is pushed to a higher energy due to the
presence of itinerant electrons, which is consistent with the previous
study \cite{Lv2010} using the Holstein-Primakoff transformation. In the paramagnetic
state, the particle-hole continuum keeps the collective spin excitation
near $\left(\pi,\pi\right)$ at a higher energy even without any $C_{4}$
symmetry breaking.

\section{Model}
The basic physics we envision being relevant to the spin-wave spectrum in the paramagnetic state is damping arising from particle-hole excitations
of the conduction electrons.  Consequently, the minimal model is the double-exchange model,
\begin{eqnarray}
\mathcal{H} & = & \mathcal{H}_{\mathrm{loc}}+\mathcal{H}_{\mathrm{itn}}+\mathcal{H}_{\mathrm{H}},
\end{eqnarray}
proposed earlier by Lv et al.\cite{Lv2010},
where $\mathcal{H}_{\mathrm{loc}}$ describes the superexchange coupling
between local moments, $\mathcal{H}_{\mathrm{itn}}$ is associated
with the itinerant electrons of the degenerate $d_{xz}$ and $d_{yz}$
orbitals, and $\mathcal{H}_{\mathrm{H}}$ describes the ferromagnetic
Hund coupling between local moments and itinerant electrons. 
The local moments are represented by a $J_{1}$-$J_{2}$ Heisenberg
model:
\begin{equation}
\mathcal{H}_{\mathrm{loc}}=\left(J_{1}\sum_{\left\langle i,j\right\rangle }+J_{2}\sum_{\left\langle \left\langle i,j\right\rangle \right\rangle }\right)\mathbf{S}_{i}\cdot\mathbf{S}_{j},
\end{equation}
where the first and second summations are performed over nearest and
next-nearest neighbors, respectively. We will focus on the $J_{1}<2J_{2}$
regime in which the system exhibits striped magnetic order. The itinerant
electrons are described by a two-band tight-binding model
\begin{eqnarray}
\mathcal{H}_{\mathrm{itn}} & = & \sum_{k\nu}\left(\begin{array}{cc}
c_{kx\nu}^{\dagger} & c_{ky\nu}^{\dagger}\end{array}\right)\left(\begin{array}{cc}
\epsilon_{k}^{x} & \epsilon_{k}^{xy}\\
\epsilon_{k}^{xy} & \epsilon_{k}^{y}
\end{array}\right)\left(\begin{array}{c}
c_{kx\nu}\\
c_{ky\nu}
\end{array}\right),
\end{eqnarray}
where 
\begin{eqnarray*}
\epsilon_{k}^{x} & = & -2t_{1}\cos k_{x}-2t_{2}\cos k_{y}-4t_{3}\cos k_{x}\cos k_{y},\\
\epsilon_{k}^{y} & = & -2t_{1}\cos k_{y}-2t_{2}\cos k_{x}-4t_{3}\cos k_{x}\cos k_{y},\\
\epsilon_{k}^{xy} & = & -4t_{4}\sin k_{x}\sin k_{y}.
\end{eqnarray*}
As defined in Ref. \onlinecite{Raghu2008,Lv2010}, the hopping parameters
$t_{1},t_{2},t_{3},t_{4}$ are between orbitals at nearest and next-nearest
neighbors. The operator $c_{i\alpha\nu}$ removes an itinerant electron
at site $i$, orbital $\alpha$ with spin $\nu$. Finally, the Hamiltonian
for the ferromagnetic Hund coupling is 
\begin{equation}
\mathcal{H}_{\mathrm{H}}=-J_{\mathrm{H}}\sum_{i\alpha}\mathbf{S}_{i}\cdot\mathbf{s}_{i\alpha},
\end{equation}
where
\begin{eqnarray*}
\mathbf{s}_{i\alpha} & = & \frac{1}{2}\sum_{\nu\nu^{\prime}}c_{i\alpha\nu}^{\dagger}\bm{\sigma}_{\nu\nu^{\prime}}c_{i\alpha\nu^{\prime}}
\end{eqnarray*}
is the spin of the itinerant electrons at site $i$ and orbital $\alpha$.

\section{Method}

\subsection{Mean-field approximation}

Based on measurements of the total fluctuating magnetic moments \cite{Gretarsson2011,Liu2012a,Harriger2012},
we assume that the local moments have spin $\frac{1}{2}$.  We then
represent the local moments as fermions using 
\begin{eqnarray}
\mathbf{S}_{i} & = & \frac{1}{2}\sum_{\nu\nu^{\prime}}f_{i\nu}^{\dagger}\bm{\sigma}_{\nu\nu^{\prime}}f_{i\nu^{\prime}},
\end{eqnarray}
and we apply a mean-field approximation to decouple the four-fermion
terms in $\mathcal{H}_{\mathrm{loc}}$ and $\mathcal{H}_{\mathrm{H}}$.
Because the system has striped magnetic order with ordering vector $Q=\left(\pi,0\right)$
at low temperatures, the staggered magnetizations, $M_{\mathrm{loc}}$
and $M_{\mathrm{itn}}$, of the local moments and itinerant electrons
are the natural mean-field order parameters. These parameters are
defined by 
\begin{eqnarray}
\left\langle S_{i}^{z}\right\rangle  & = & M_{\mathrm{loc}}e^{iQ\cdot r_{i}},\\
\sum_{\alpha}\left\langle s_{i\alpha}^{z}\right\rangle  & = & M_{\mathrm{itn}}e^{iQ\cdot r_{i}}.
\end{eqnarray}
In addition, we fix the expectation values of the nearest and next-nearest
neighbor exchange terms $\chi_{1},\chi_{2}=\frac{1}{N}\sum_{i}\left\langle f_{i}^{\dagger}f_{j}\right\rangle $
at non-zero values, so that the mean-field Hamiltonian in the paramagnetic
state does not vanish. Such non-zero $\chi_{1},\chi_{2}$ can be obtained from the Hubbard model from which a Heisenberg model is typically derived.

The mean-field Hamiltonian for the local moment is 
\begin{eqnarray}
\mathcal{H}_{\mathrm{loc}}^{\mathrm{MF}} & = & \sum_{k\sigma}\left(A_{k\sigma}f_{k\sigma}^{\dagger}f_{k\sigma}+B_{k\sigma}f_{k\sigma}^{\dagger}f_{k+Q,\sigma}\right),
\end{eqnarray}
where 
\begin{eqnarray*}
A_{k\sigma} & = & -\frac{3}{4}J_{1}\chi_{1}\left(\cos k_{x}+\cos k_{y}\right)-\frac{3}{2}J_{2}\chi_{2}\cos k_{x}\cos k_{y},\\
B_{k\sigma} & = & -2J_{2}M_{\mathrm{loc}}\sigma.
\end{eqnarray*}
Here, $\sigma=\pm1$ corresponds to up and down spins, respectively.
Similarly, the mean-field Hamiltonian for the Hund coupling is 
\begin{eqnarray}
\mathcal{H}_{\mathrm{H}}^{\mathrm{MF}} & \equiv & \delta\mathcal{H}_{\mathrm{itn}}^{\mathrm{MF}}+\delta\mathcal{H}_{\mathrm{loc}}^{\mathrm{MF}},
\end{eqnarray}
where 
\begin{eqnarray*}
\delta\mathcal{H}_{\mathrm{itn}}^{\mathrm{MF}} & = & -\frac{1}{2}J_{\mathrm{H}}M_{\mathrm{loc}}\sum_{k\alpha\nu}\nu c_{k\alpha\nu}^{\dagger}c_{k+Q,\alpha\nu},\\
\delta\mathcal{H}_{\mathrm{loc}}^{\mathrm{MF}} & = & -\frac{1}{2}J_{\mathrm{H}}M_{\mathrm{itn}}\sum_{k\nu}\nu f_{k\nu}^{\dagger}f_{k+Q,\nu}.
\end{eqnarray*}
Hence, at the mean-field level, the itinerant electrons and local
moments are decoupled, and they are effectively governed by $\mathcal{H}_{\mathrm{itn}}^{\mathrm{eff}}=\mathcal{H}_{\mathrm{itn}}+\delta\mathcal{H}_{\mathrm{itn}}^{\mathrm{MF}}$
and $\mathcal{H}_{\mathrm{loc}}^{\mathrm{eff}}=\mathcal{H}_{\mathrm{loc}}^{\mathrm{MF}}+\delta\mathcal{H}_{\mathrm{loc}}^{\mathrm{MF}}$,
respectively. 

For convenience, we introduce the three-component operator $c_{ka\nu}=\left(c_{kx\nu},c_{ky\nu},f_{k\nu}\right)$,
for $a=1,2,3$. Then, the full mean-field Hamiltonian $\mathcal{H}^{\mathrm{MF}}=\mathcal{H}_{\mathrm{loc}}^{\mathrm{eff}}+\mathcal{H}_{\mathrm{itn}}^{\mathrm{eff}}$
can be diagonalized by unitary transformations 
\begin{eqnarray}
c_{ka\nu} & = & \sum_{n=1}^{6}U_{k\nu,1an}d_{k\nu n},\\
c_{k+Q,a\nu} & = & \sum_{n=1}^{6}U_{k\nu,2an}d_{k\nu n},
\end{eqnarray}
for $k$ in the reduced Brillouin zone, to give $\mathcal{H}^{\mathrm{MF}}=\sum_{k\nu n}^{\prime}E_{k\nu n}d_{k\nu n}^{\dagger}d_{k\nu n}$.
The prime over the summation indicates a $k$-summation over the reduced
Brillouin zone.  The mean-field order parameters $M_{\mathrm{loc}}$ and $M_{\mathrm{itn}}$
can then be found by solving the self-consistent equations 

\begin{eqnarray}
M_{\mathrm{loc}} & = & \frac{1}{N}\sum_{k\nu m}^{\prime}\nu U_{k\nu,13m}U_{k\nu,23m}n_{k\nu m},\\
M_{\mathrm{itn}} & = & \frac{1}{N}\sum_{k\nu m}^{\prime}\sum_{a=1}^{2}\nu U_{k\nu,1am}U_{k\nu,2am}n_{k\nu m},
\end{eqnarray}
where $n_{k\nu m}=\left\langle d_{k\nu m}^{\dagger}d_{k\nu m}\right\rangle $
is the Fermi-Dirac occupancy number of the diagonalized bands.

\subsection{Dynamic spin susceptibility}

The transverse spin susceptibility of the system is given by the correlation
function between the various spin operators. Since there are three
species of fermions, the spin susceptibility is a $3\times3$ matrix,
\begin{eqnarray}
\chi_{0,ab}^{+-}\left(q,q^{\prime};t\right) & = & -i\theta\left(t\right)\left\langle \left[S_{q,a}^{+}\left(t\right),S_{-q^{\prime},b}^{-}\left(0\right)\right]\right\rangle ,
\end{eqnarray}
where $S_{q,a}$ is the spin operator corresponding to $c_{qa\nu}$.
Because of the doubling of the unit cell in the ordered state, the susceptibility,
\begin{eqnarray}
\chi_{0,ab}^{+-}\left(q,q^{\prime},\omega\right) & = & \frac{1}{N}\sum_{kmm^{\prime}}^{\prime}\frac{n_{k\uparrow m}-n_{k+q,\downarrow m^{\prime}}}{\omega+E_{k\uparrow m}-E_{k+q,\downarrow m^{\prime}}+i\delta}\nonumber \\
 &  & \qquad\times\gamma_{qak,mm^{\prime}}\gamma_{q^{\prime}bk,mm^{\prime}}^{*}\nonumber \\
 &  & \qquad\times\left(\delta_{q,q^{\prime}}+\delta_{q,q^{\prime}+Q}\right),
\end{eqnarray}
is non-zero for $q=q^{\prime}$ and $q=q^{\prime}+Q$, 
where
\begin{eqnarray*}
\gamma_{qak,mm^{\prime}} & = & \sum_{\xi=1}^{2}U_{k\uparrow,\xi am}^{*}U_{k+q,\downarrow,\tau\left(k+\xi,Q+q\right),am^{\prime}}.
\end{eqnarray*}
Here, $\tau\left(k\right)$ equals $2$ for $k$ in the reduced Brillouin
zone, and equals $1$ otherwise. The system size is denoted by $N$,
and a small positive $\delta$ is included for convergence. 

To include the interaction effects, we apply a generalized random
phase approximation. The resulting susceptibility $\bar{\chi}^{+-}\left(q,q^{\prime};\omega\right)$
is given by the Dyson equation
\begin{eqnarray}
\bar{\chi}^{+-}\left(q,q^{\prime};\omega\right) & = & \chi_{0}^{+-}\left(q,q^{\prime};\omega\right)\nonumber \\
 &  & +\sum_{q^{\prime\prime}}\chi_{0}^{+-}\left(q,q^{\prime\prime};\omega\right)U_{q^{\prime\prime}}\bar{\chi}^{+-}\left(q^{\prime\prime},q^{\prime};\omega\right),\nonumber \\
\end{eqnarray}
where the non-zero entries of the interaction matrix $U_{q}$ are
$U_{q,13}=U_{q,23}=U_{q,31}=U_{q,32}=-\frac{1}{2}J_{\mathrm{H}}$,
and $U_{q,33}=J_{1}\left(\cos q_{x}+\cos q_{y}\right)+2J_{2}\cos q_{x}\cos q_{y}$.
It is straightforward to show that the solution has the form $\bar{\chi}^{+-}=\left[I-\chi_{0}^{+-}U\right]^{-1}\chi_{0}^{+-}.$
The quantity to be compared with the INS measurements is the total spin
susceptibility $\bar{\chi}_{\mathrm{tot}}^{+-}\left(q,\omega\right)$,
defined as the sum of all $3\times3$ components of $-\bar{\chi}^{+-}\left(q,q;\omega\right)$.

\section{Results}

\subsection{Mean-field approximation}

For modeling purposes, we set $J_{1}=0.16$ and $J_{2}=0.6J_{1}$ as in Ref. \onlinecite{Lv2010}.
However, we choose from Ref. \onlinecite{Raghu2008} an alternate
set of tight-binding parameters, because these parameters more accurately
reproduce the Fermi surfaces found in angle-resolved photoemission
spectroscopy (ARPES) experiments \cite{Ding2008} and first-principles band structure calculations \cite{Cao2008}. Explicitly, we set $t_{1}=-0.5$, $t_{2}=0.65$,
and $t_{3}=t_{4}=-0.425$, which gives a bandwidth comparable to that
in Ref. \onlinecite{Lv2010}. Finally, we also set $J_{H}=4$, $\chi_{1}=\chi_{2}=0.2$,
and we fix the filling of the itinerant bands at $n=2.1$. 

\begin{figure}[H]
\centering{}\includegraphics[width=0.95\columnwidth]{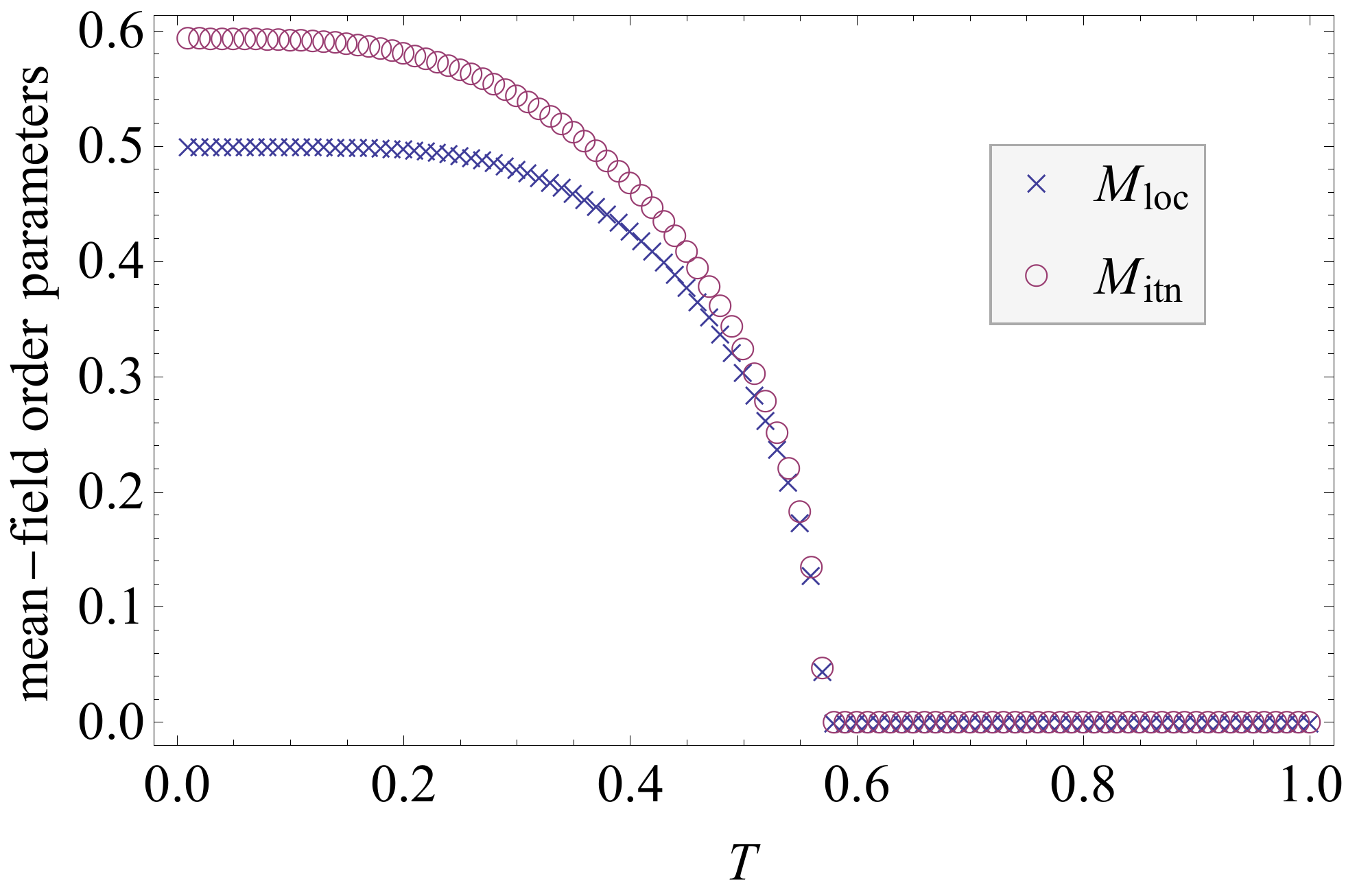}\caption{The temperature dependence of the mean-field order parameters $M_{\mathrm{loc}}$
and $M_{\mathrm{itn}}$. \label{fig:order_param_vs_T}}
\end{figure}

Figure \ref{fig:order_param_vs_T} shows the temperature dependence
of the order parameters. The local moment magnetization $M_{\mathrm{loc}}$
saturates at a value of $0.5$, while the itinerant electron magnetization
$M_{\mathrm{itn}}$ saturates at a value that depends on the Hund
coupling and the filling of the itinerant bands. In addition, both
the local moments and itinerant electrons have the same transition
temperature. While a model incorporating $\mathcal{H}_{\mathrm{itn}}$
alone does not order magnetically, the inclusion of the Hund coupling
term $\mathcal{H}_{\mathrm{H}}$  imposes on the itinerant electrons the striped magnetic order of the local moments. 
While the mean-field approximation is not expected to yield an accurate
value for the transition temperature, we note that the Hund coupling increases
the transition temperature. This implies that the presence of the
itinerant electrons stabilizes the magnetic order of the local moments.

\begin{figure}[H]
\begin{centering}
\includegraphics[width=0.95\columnwidth]{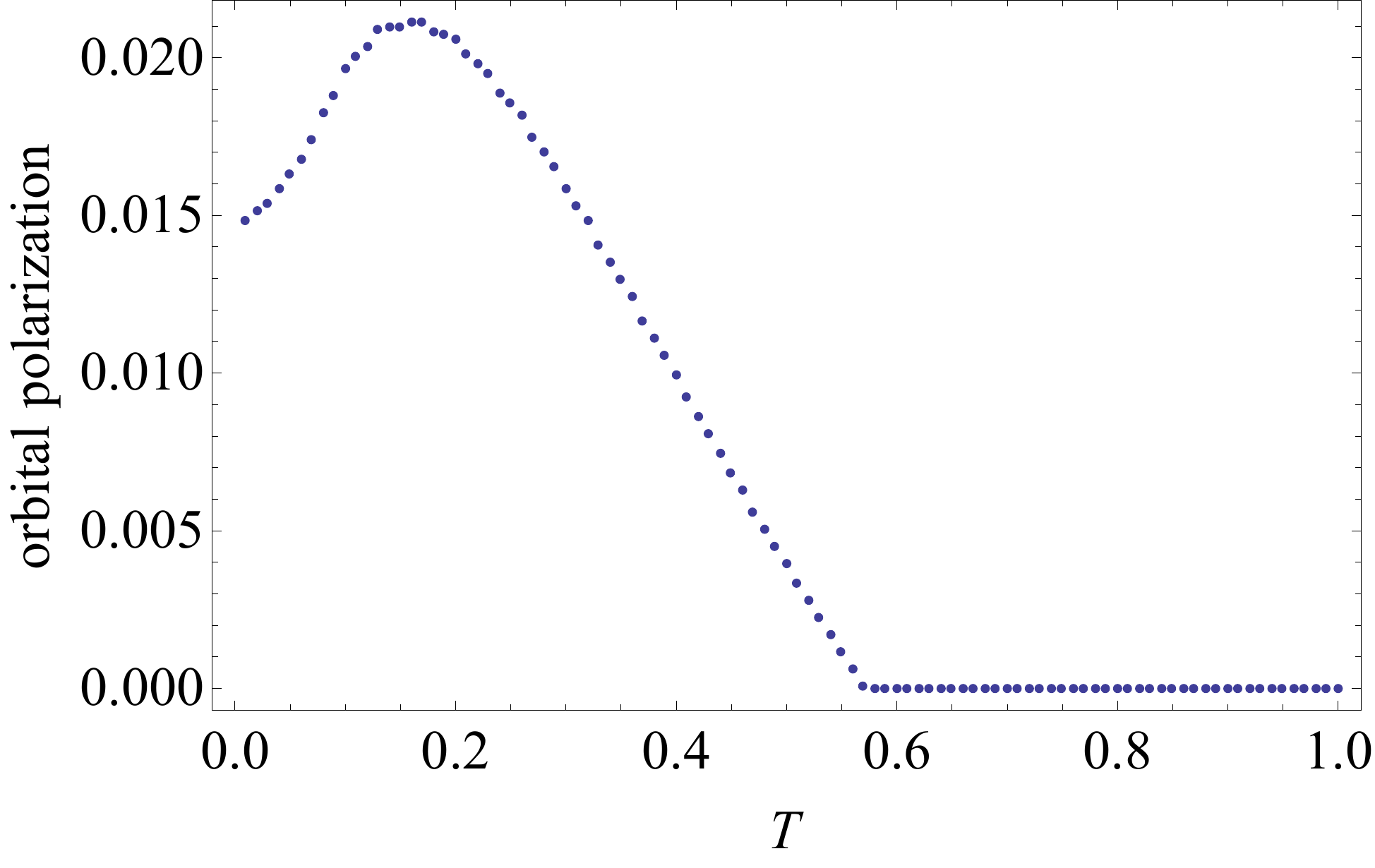}
\par\end{centering}

\caption{The temperature dependence of the orbital polarization. \label{fig:orbital_polarization}}
\end{figure}

As discussed in Ref. \onlinecite{Lv2010}, the degeneracy between
the $d_{xz}$ and $d_{yz}$ orbitals is broken in the ordered state
by the Hund coupling. Such an orbital ordering was observed in ARPES
measurements \cite{Yi2011}.  Figure \ref{fig:orbital_polarization}
shows the temperature dependence of the orbital polarization, defined
as the occupancy difference between the $d_{xz}$ and $d_{yz}$ orbitals.
As the temperature increases, the orbital polarization decreases, vanishing
at the same temperature as the mean-field order parameters. The increase
at low temperature is not a general feature, and can be accounted
for by considering the details of the itinerant bands.

\subsection{Dynamic spin susceptibility}

\begin{figure}[H]
\begin{centering}
\subfloat[\label{fig:suscep_low_T}]{\begin{centering}
\includegraphics[width=0.95\columnwidth]{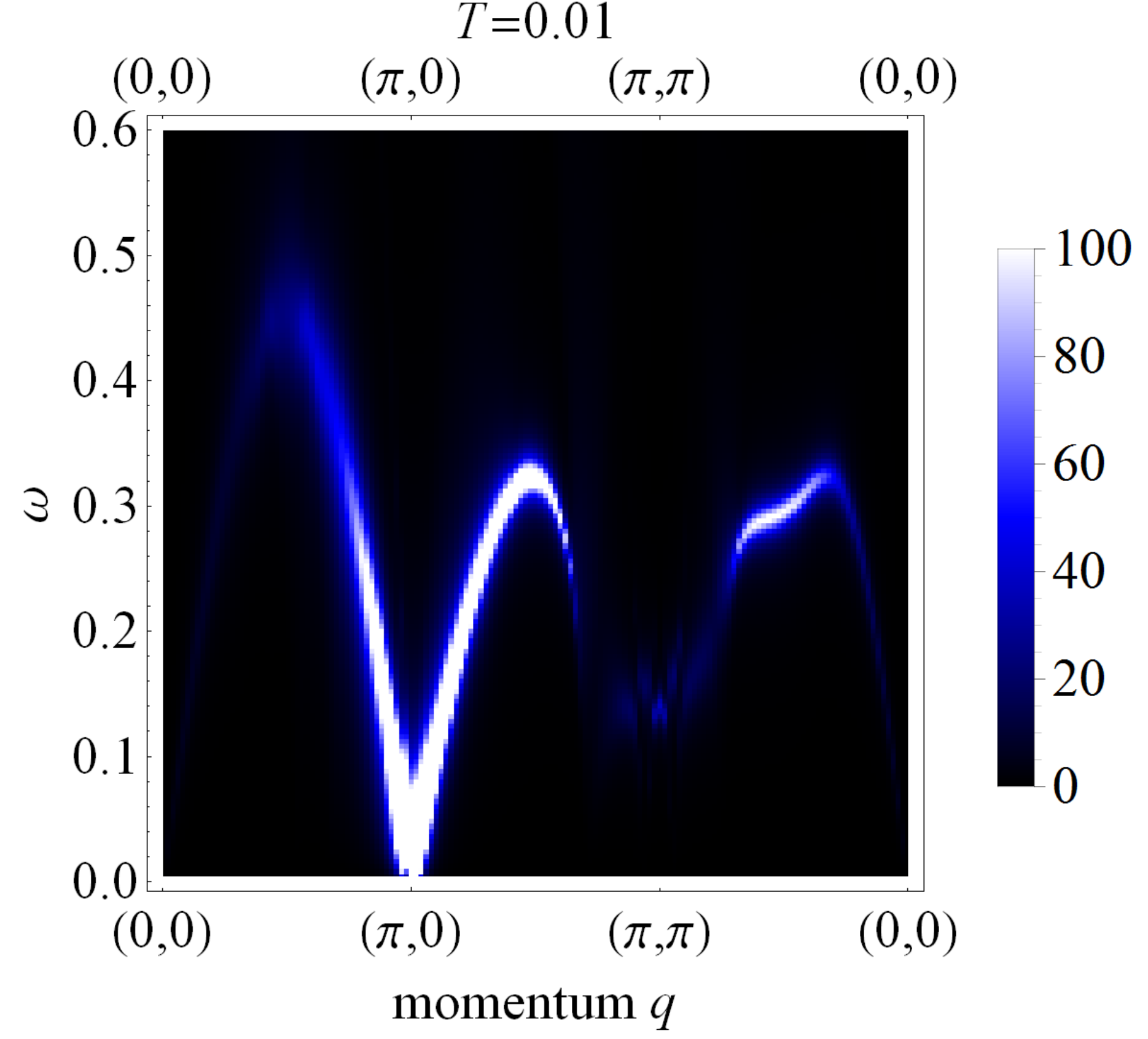}
\par\end{centering}

}
\par\end{centering}

\begin{centering}
\subfloat[]{\begin{centering}
\includegraphics[width=0.95\columnwidth]{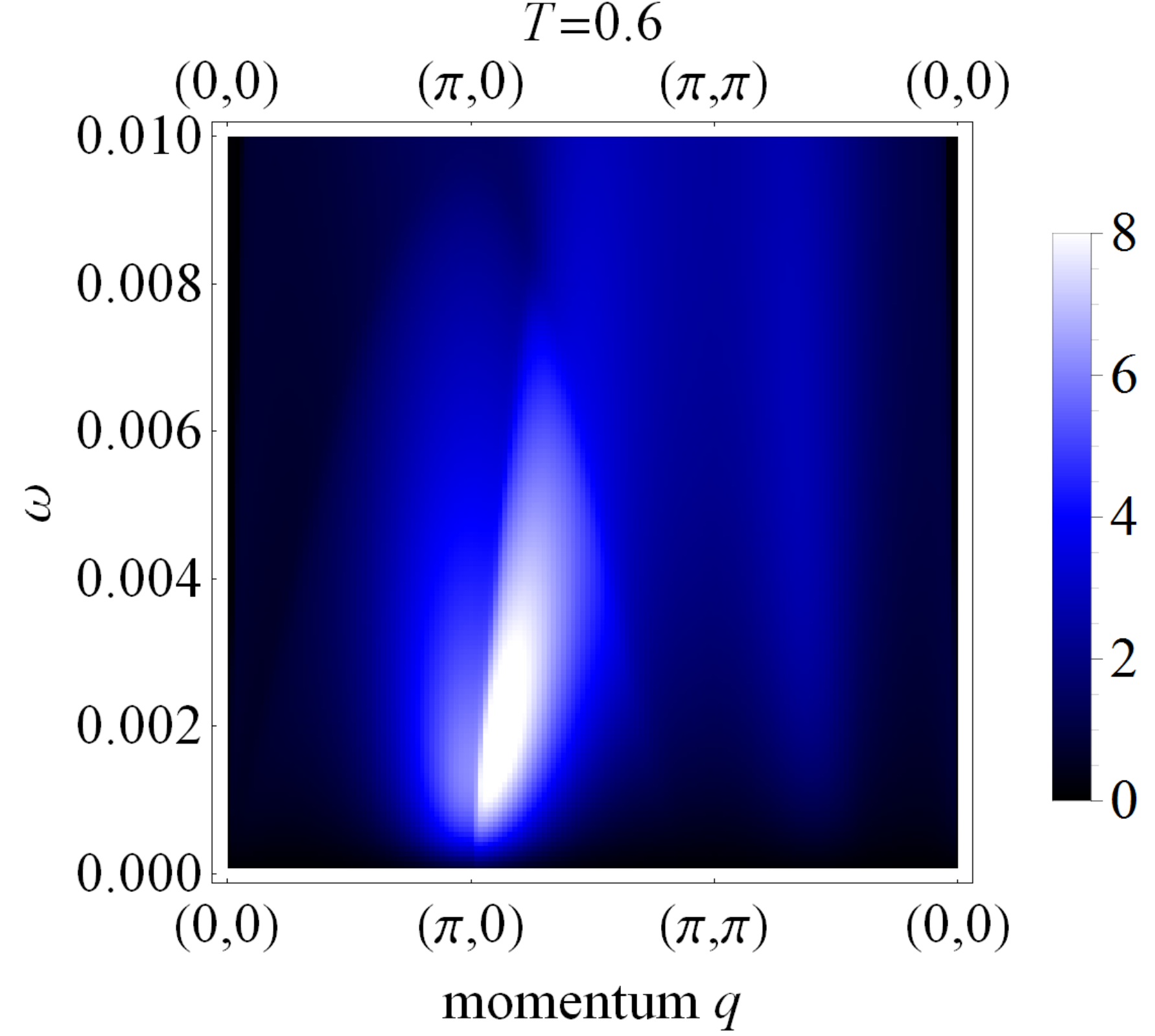}
\par\end{centering}

}
\par\end{centering}

\caption{The total spin susceptibility $\mbox{Im}\bar{\chi}_{\mathrm{tot}}^{+-}$
of the degenerate double-exchange model along the path $\left(0,0\right)$-$\left(\pi,0\right)$-$\left(\pi,\pi\right)$-$\left(0,0\right)$
in the (a) ordered and (b) paramagnetic state. The tight-binding parameters
are $t_{1}=-0.5$, $t_{2}=0.65$, and $t_{3}=t_{4}=-0.425$, with
an itinerant band filling of $n=2.1$. The superexchange couplings
are $J_{1}=0.16$ and $J_{2}=0.6J_{1}$. The Hund coupling is $J_{H}=4$.
\label{fig:suscep_vary_T}}
\end{figure}

\begin{figure}[H]
\begin{centering}
\subfloat[\label{fig:bare_suscep_low_T}]{\begin{centering}
\includegraphics[width=0.95\columnwidth]{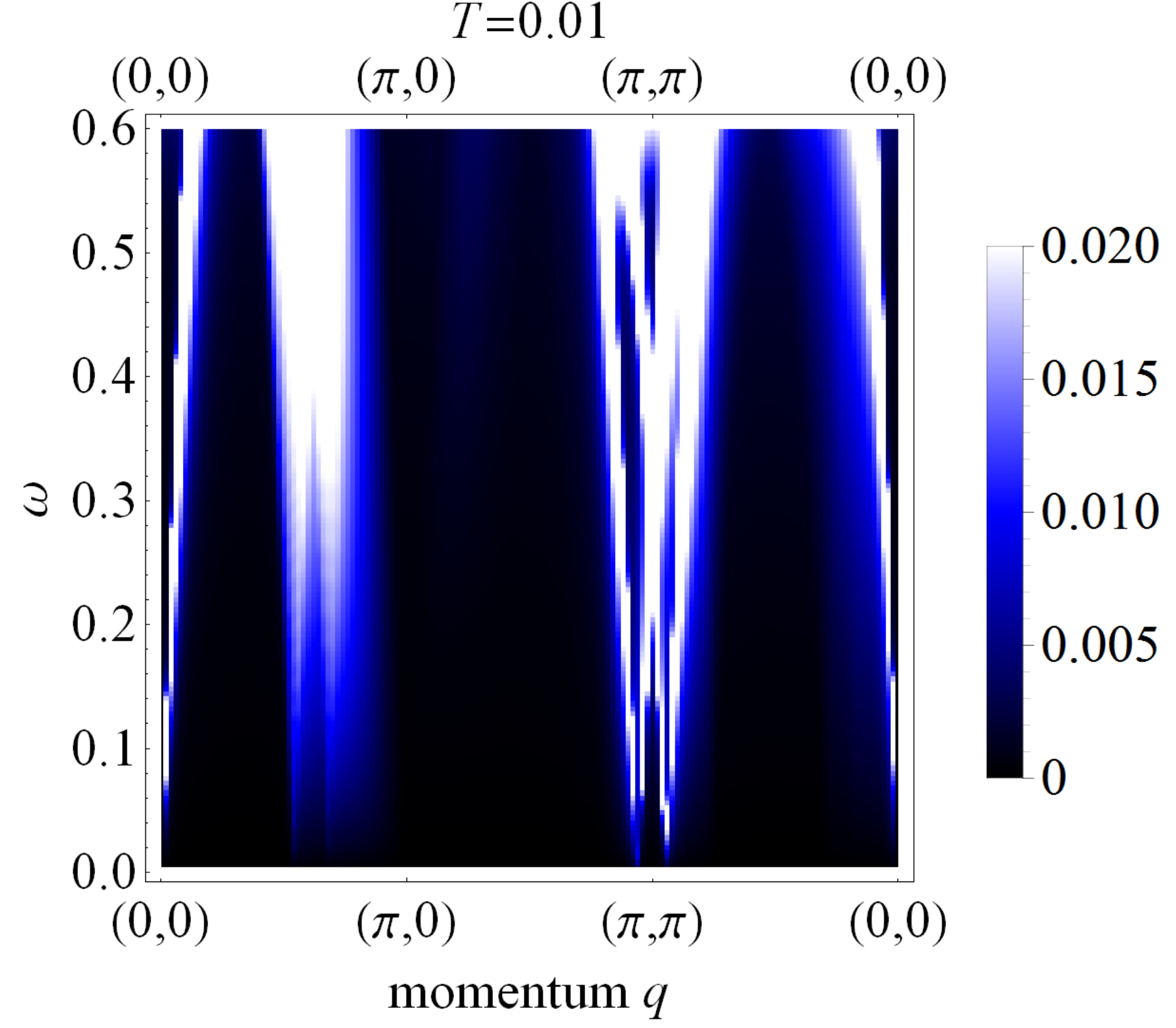}
\par\end{centering}

}
\par\end{centering}

\begin{centering}
\subfloat[\label{fig:bare_suscep_high_T}]{\begin{centering}
\includegraphics[width=0.95\columnwidth]{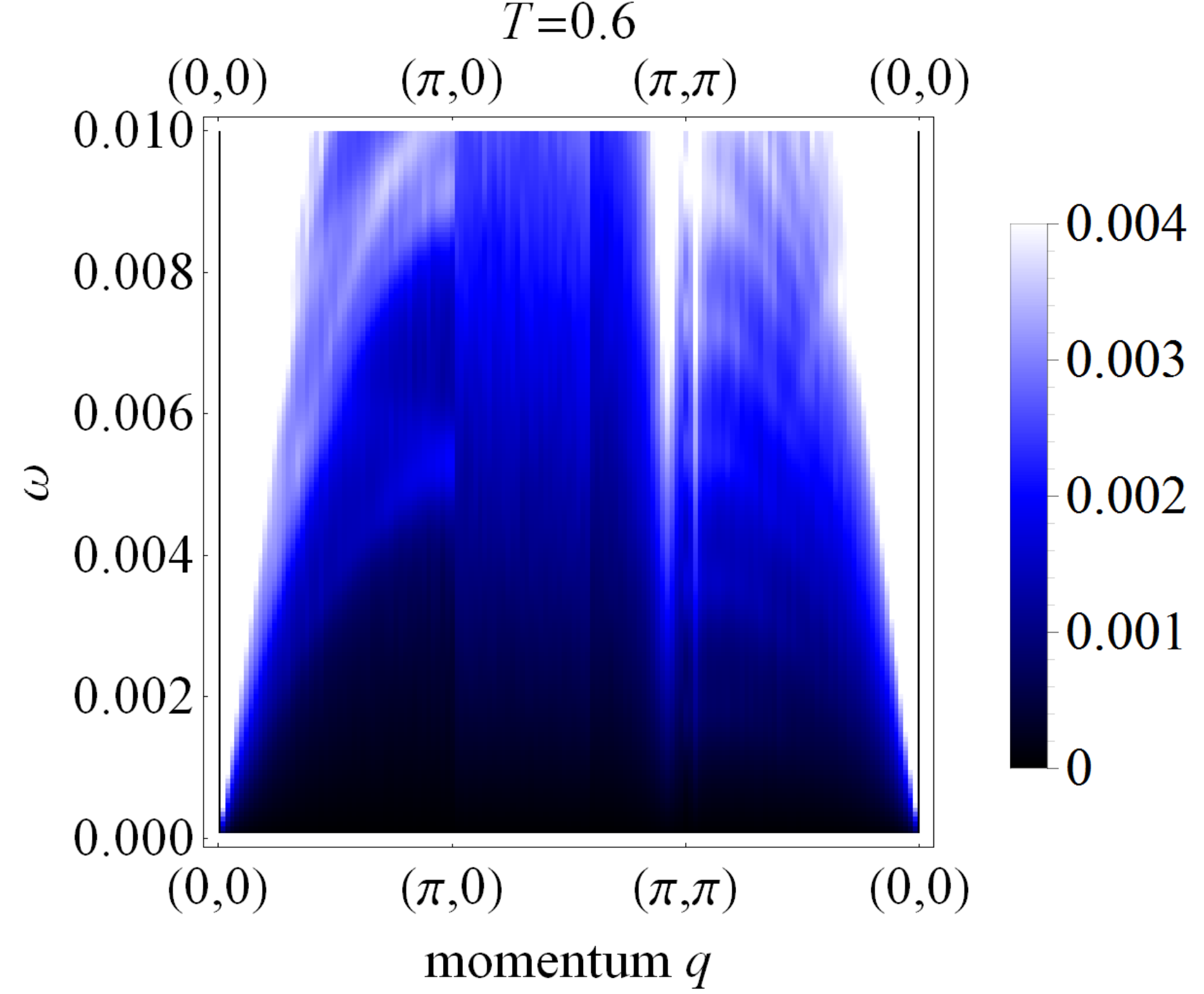}
\par\end{centering}

}
\par\end{centering}

\caption{The itinerant components of the bare susceptibility $\mbox{Im} \chi_0^{+-} $ in the (a) ordered and (b) paramagnetic state. The particle-hole continuum dampens the spin-wave excitations.
\label{fig:bare_suscep}}
\end{figure}

For numerical purposes, we use a system size of $N=1000\times1000$
and $\delta=0.01$ for the ordered state, and $\delta=0.0005$ for
the paramagnetic state. A smaller $\delta$ is used for the paramagnetic state so that the energy resolution is appropriate for the lower energy scale involved. A larger $N$ and a smaller $\delta$ do not
change our results qualitatively. Figure \ref{fig:suscep_vary_T}
shows  both the imaginary
part of the total spin susceptibility for the momentum-space path
$\left(0,0\right)$-$\left(\pi,0\right)$-$\left(\pi,\pi\right)$-$\left(0,0\right)$ in both the (a) ordered and (b) paramagnetic states.
In the ordered state, the Hund coupling raises the excitation energy
at $\left(\pi,\pi\right)$. This effect can be attributed to orbital
ordering \cite{Lv2010}, which stabilizes order at $\left(\pi,0\right)$
at the expense of competing order at $\left(\pi,\pi\right)$. While
the orbital polarization here is an order of magnitude smaller than
that found in Ref. \onlinecite{Lv2010}, the effect at $\left(\pi,\pi\right)$
remains significant. In addition, the presence of itinerant electrons
dampens the excitations around $\left(\pi,\pi\right)$. Figure \ref{fig:bare_suscep_low_T} shows the itinerant components of the bare spin susceptibility $\mbox{Im}\chi_0^{+-}$. The regions with strong particle-hole continuum correspond to regions with heavily damped spin-wave excitations. These observations
are consistent with the results of INS measurements \cite{Diallo2009,Zhao2009,Ewings2011,Harriger2011,Harriger2012}.

Figure \ref{fig:suscep_low_T_(pi,pi)} shows the temperature dependence
of the spin susceptibility along $\left(\pi,\pi\right)$. As the temperature
increases, while the excitations around $\left(\pi,\pi\right)$ soften, the Landau
damping in the same region increases. Further experiments will be necessary to verify this feature.  In the paramagnetic state, the strong spin-wave-like excitation near
$\left(\pi,0\right)$ persists, while the particle-hole continuum shown in Figure \ref{fig:bare_suscep_high_T}
pushes any collective spin excitations near $\left(\pi,\pi\right)$
to a higher energy. This feature is robust, because a finite particle-hole
continuum always exists at the finite wavevector $\left(\pi,\pi\right)$,
provided that the single-particle energy spectrum is not fully gapped.
Unlike previous theoretical models \cite{Goswami2011,Yu2012}, our
results are obtained without breaking $C_{4}$ symmetry. This makes
our results applicable to INS measurements even at high temperatures
\cite{Harriger2012}. This is the key finding of this work.

In our calculation, as the temperature increases in the ordered state,
the energy scale of the collective excitations decreases together
with the mean-field order parameters. In the paramagnetic state, the
energy scale is simply fixed by $\chi_{1},\chi_{2}$. These observations
are inconsistent with INS measurements, which show that the energy
scale of the collective excitations is independent of temperature.
This inconsistency likely arises from the limitations of the mean-field
approximation.

\begin{figure}[H]
\begin{centering}
\includegraphics[width=0.95\columnwidth]{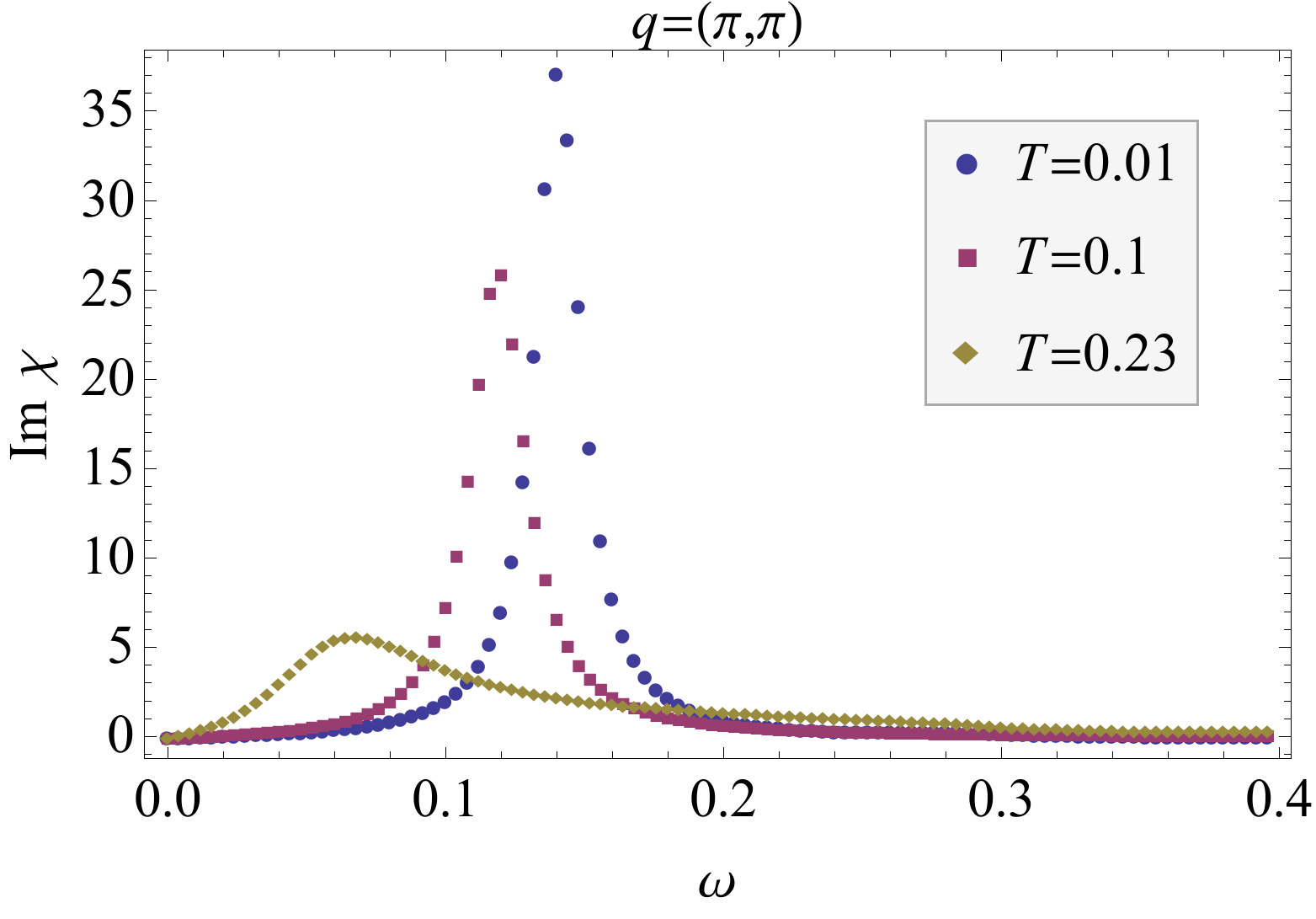}
\par\end{centering}

\caption{The spin susceptibility along the wavevector $\left(\pi,\pi\right)$
at various temperatures. As temperature increases, excitations around
$\left(\pi,\pi\right)$ soften and become more damped.\label{fig:suscep_low_T_(pi,pi)}}
\end{figure}

\section{Discussion and Conclusion}

We also calculated the low-temperature spin susceptibility (not shown)
using the tight-binding parameters in Ref. \onlinecite{Lv2010}. The
excitation energy spectrum is consistent with previous results obtained
using the Holstein-Primakoff representation and a linear spin-wave
approximation. Compared to the spectrum in Figure \ref{fig:suscep_low_T},
the excitation energy at $\left(\pi,\pi\right)$ has a larger increase
due to a stronger orbital order, but the excitations around $\left(\pi,\pi\right)$
are less damped. Therefore, while our results do not qualitatively
depend on the choice of parameters, different parameters can be used
to produce the quantitative differences between various types of iron
pnictides. Furthermore, the $\left(\pi,0\right)$-ordering is robust
because the paramagnetic spin susceptibility exhibits a peak at $\left(\pi,0\right)$
despite the itinerant bands having imperfect Fermi surface nesting.
This is in contrast with the calculations using only the itinerant model in Ref. \onlinecite{Kaneshita2010},
which show incommensurate peaks. 

Our results at high temperatures are consistent with first-principles
calculations based on a combination of density functional theory and
dynamical mean-field theory \cite{Park2011}. This suggests that our
model has captured the essential physics of spin excitations in iron
pnictides. Since the mechanism for superconductivity is believed to
arise from spin fluctuations, it would be important to consider both
the local moments and itinerant electrons when studying superconductivity
in iron pnictides.

For our choice of parameters, the $d_{xz}$ orbital has a larger occupancy
than the $d_{yz}$ orbital. This is opposite the result obtained in
Ref.  \onlinecite{Lv2010}. This difference arises because the opposite sign between
the two sets of tight-binding parameters makes occupying the $d_{xz}$ orbital more energetically favorable. This higher occupancy of the $d_{xz}$ orbital agrees with ARPES measurements \cite{Yi2011}, 
which show that the $d_{xz}$ orbital is lower in energy than the $d_{yz}$ orbital in the magnetically ordered state.

To close,  we studied the spin excitation spectra of the degenerate
double-exchange model. This model consists of local moments represented
by a $J_{1}$-$J_{2}$ Heisenberg model, and itinerant electrons from
the degenerate $d_{xz}$ and $d_{yz}$ orbitals represented by a tight-binding
model. The local moments and itinerant electrons are coupled through
a ferromagnetic Hund coupling. Using a fermionic representation of
the local moments and a generalized random phase approximation, we
obtained a unified framework for the spin excitations in both the
ordered and paramagnetic state. The calculated spin susceptibility
shows energy spectra and Landau damping consistent with measurements
from inelastic neutron scattering experiments over a wide range of
temperatures.
\begin{acknowledgments}
We thank J. Knolle for an email exchange which led to our inclusion of Fig. 4.  Z. Leong is supported by a scholarship from the Agency of Science, Technology and Research. W. Lv is supported by NSF Grant No. DMR-1104386.  W. C. Lee and P. Phillips are supported by the Center for Emergent Superconductivity, a DOE Energy Frontier Research Center, Grant No. DE-AC0298CH1088.
\end{acknowledgments}
\bibliographystyle{apsrev4-1}
\bibliography{library}

\end{document}